\documentclass[journal=jpclcd,manuscript=letter]{achemso}
\SectionNumbersOn
\usepackage[version=3]{mhchem} 
\usepackage{rotating}

\usepackage{amssymb, amsmath}
\usepackage[english]{babel}
\usepackage{bm}
\usepackage{graphicx}
\usepackage[latin1]{inputenc}
\usepackage{color}
\usepackage{xr}
\newcommand{\RANGLE}%
{\mathchoice{\bigr\rangle}{\bigr\rangle}{\rangle}{\rangle}}
\newcommand{\LANGLE}%
{\mathchoice{\bigl\langle}{\bigl\langle}{\langle}{\langle}}

\author{Konstantin E. Dorfman}
\email{kdorfman@uci.edu}

\author{Frank Schlawin}

\author{Shaul Mukamel}

\affiliation{Department of Chemistry, University of California, Irvine,
California 92697-2025, USA}
\date{\today}%

\title{Stimulated Raman Spectroscopy with Entangled Light;\\ Enhanced Resolution and Pathway Selection}

\begin{document}

\begin{abstract}
We propose a novel femtosecond stimulated Raman spectroscopy (FSRS) technique that combines entangled photons with interference detection to select matter pathways and enhance the resolution. Following photo excitation by an actinic pump, the measurement uses a pair of broadband entangled photons, one (signal) interacts with the molecule  together with a third narrowband pulse induces the Raman process. The other (idler) photon provides a reference for the coincidence measurement. This interferometric photon-coincidence counting detection allows to separately measure Raman  gain and  loss signals, which is not possible with conventional probe  transmission detection. Entangled photons further provide a unique temporal and spectral detection window that can better resolve fast excited state dynamics compared to classical and correlated disentangled states of light.
\end{abstract}

\paragraph*{Keywords:} Femtosecond Raman Spectroscopy, entangled photons, interferometry

\clearpage

Stimulated Raman spectroscopy is one of the most versatile tools for the study of molecular vibrations. Applications include probing time-resolved photophysical and photochemical processes \cite{Kukura:AnnurevPhysChem:2007,Kuk05,Tak08,Kuramochi:JPCLett:2012}, chemically specific biomedical imaging \cite{Xin02}, and chemical sensing\cite{Aro12,Bre13}. Considerable effort has been devoted eliminate off-resonant background thus improving the signal to noise ratio and the ability to detect small samples and even single molecules. Pulse shaping \cite{Oro02,Pes07} and the combination of broad and narrow band pulses (technique known as Femtosecond Stimulated Raman Spectroscopy (FSRS) \cite{Kukura:AnnurevPhysChem:2007}) were employed. Recent measurements of absorption spectra with entangled photons in an interferometric setup\cite{Sca03,Yab04,Kal07,Sla13,Kal14} suggest a possibility to use more elaborate detection. Here we propose an Interferometric FSRS (IFSRS)  technique that combines quantum entangled light with interferometric detection to significantly enhance the resolution  and selectivity of Raman signals. By counting photons,  IFSRS can further measure separately the gain and loss contributions to the Raman spectra \cite{Har13} which is not possible with classical FSRS.

Entangled light is widely used in quantum information \cite{Lvo09,Wal11}, secure communication \cite{Gis02} and quantum computing \cite{Kok07} applications. It has been demonstrated that the twin photon state may be used to manipulate  two photon absorption $\omega_1+\omega_2$ type resonances in aggregates \cite{Sal98,Lee06, Sch13,Dor14} but these ideas do not apply to Raman $\omega_1-\omega_2$ resonances. We show that this can be achieved by using interferometric photon coincidence detection, which further enhances the signal-to-noise ratio. Moreover, entangled two-photon absorption has also been shown experimentally to scale  linearly rather than quadratically with the pump intensity\cite{Jav90,Lee06} thus allowing to use very weak light intensities, limiting damage and overcoming the photodetector noise when employing the photon coincidence measurement \cite{Bri06}.

\begin{figure*}[t]
\begin{center}
\includegraphics[trim=0cm 0cm 0cm 0cm,angle=0, width=0.95\textwidth]{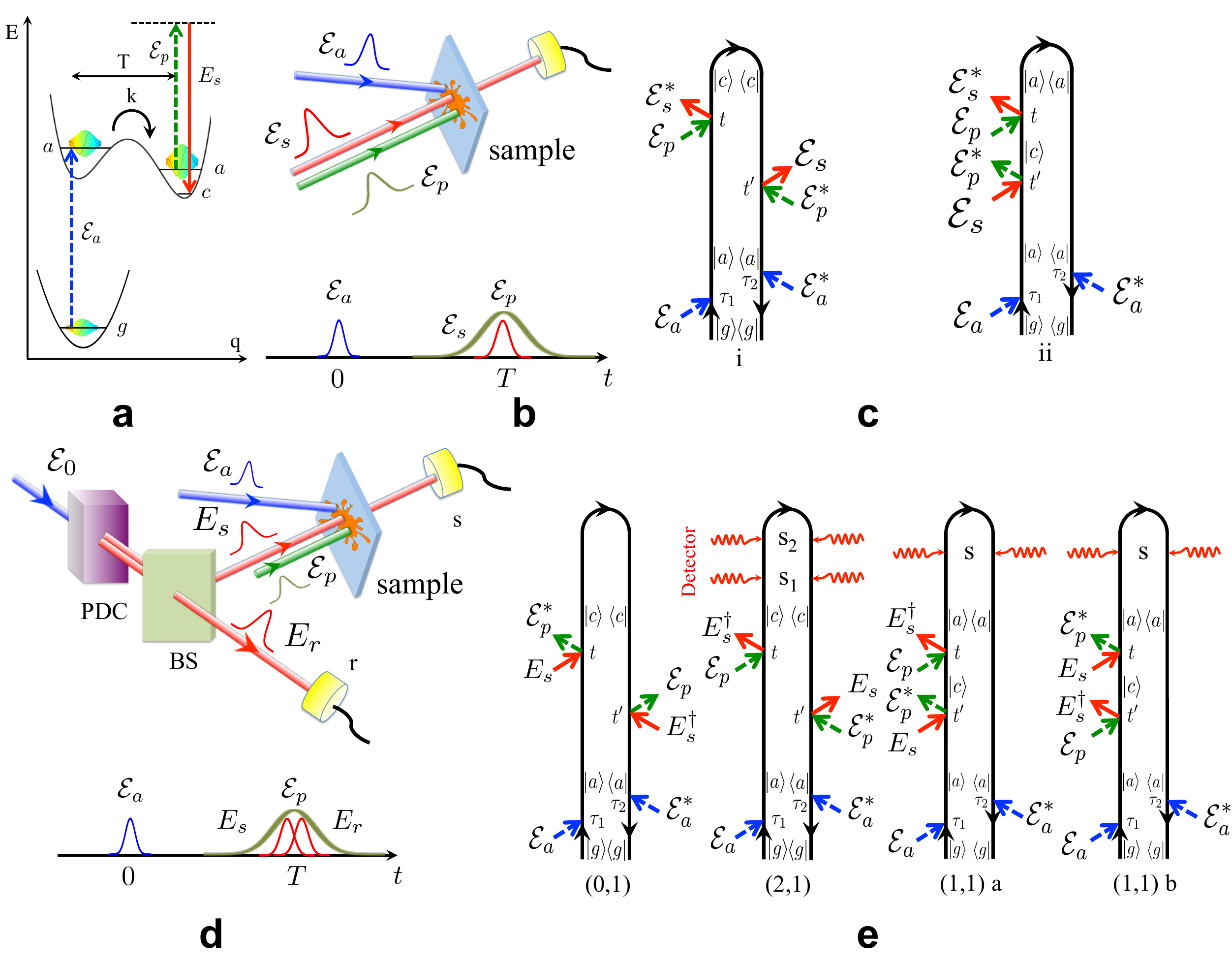}
\end{center}
\caption{(Color online) Top row: classical FSRS level scheme for the tunneling model - $\textbf{a}$, pulse configuration  - $\textbf{b}$, and  loop diagrams (for diagram rules see \cite{Rah10}) for classical FSRS - $\textbf{c}$. $\textbf{d}$ and $\textbf{e}$ - the same as $\textbf{b}$, and $\textbf{c}$ but for IFSRS. BS  and PDC in $\textbf{d}$ are the beam splitter and parametric down conversion, respectively. The pairs of indices  $(0,1)$ etc. in $\textbf{e}$ indicate number of photons registered by detectors $s$ and $r$ in each photon counting signal:  $(N_s,N_r)$}
\label{fig:setup}
\end{figure*}

In conventional FSRS, an actinic resonant pulse $\mathcal{E}_a$ first creates a vibrational wave packet in an electronically excited state (see  \ref{fig:setup}a,b). After a delay $T$, the frequency resolved transmission of a broadband (femtosecond) probe $E_s$ in the presence of a narrowband (picosecond) pump $\mathcal{E}_p$ shows excited state vibrational resonances  generated by an off-resonant stimulated Raman process.  The FSRS signal is given by \cite{Dor13}
\begin{align}\label{eq:Sc0}
&S_{FSRS}(\omega,T)=\frac{2}{\hbar}\mathcal{I}\int_{-\infty}^{\infty}dte^{i\omega(t-T)}\mathcal{T}\mathcal{E}_s^{*}(\omega)\mathcal{E}_p(t)\langle\alpha(t)e^{-\frac{i}{\hbar}\int H'_-(\tau)d\tau}\rangle,
\end{align}
where $\alpha$ is the electronic polarizability, $\mathcal{I}$ denotes the imaginary part, $\langle...\rangle=\text{tr}[...\rho]$ with $\rho$ being the density operator of the entire system, and $\mathcal{E}_s=\langle E_s\rangle$ is expectation value of the probe field operator with respect to classical state of light (hereafter $\mathcal{E}$  denotes classical fields  and $E$ stands for quantum fields). $H_-'$ is the Hamiltonian superoperator\cite{Har08} in the interaction picture (see Section \ref{sec:tf} of SI).  The exponent in Eq. (\ref{eq:Sc0}) can be expanded  perturbatively in field-matter interactions (see Section \ref{sec:FSRS} of SI). Off-resonance Raman processes can be described by the radiation/matter interaction Hamiltonian
$H'(t)=\alpha E_s^{\dagger}(t)\mathcal{E}_p(t)+ \mathcal{E}_a^{*}(t)V+H.c.$, where  $V$ is the dipole moment, $\alpha$ is the off-resonant polarizability. In the present applications we expand the signal (\ref{eq:Sc0}) to sixth order in the fields $\sim \mathcal{E}_s^2\mathcal{E}_p^2\mathcal{E}_a^2$. The resulting classical FSRS signal is given by the two diagrams in \ref{fig:setup}c which translates into Eqs. (\ref{eq:Sisr1}) - (\ref{eq:Siisr1}) of SI.
All relevant matter information is contained in the two four-point correlation functions 
\begin{align}\label{eq:Fi}
F_i(t_1,t_2,t_3)=\langle V G^{\dagger}(t_1)\alpha G^{\dagger}(t_2)\alpha G(t_3)V^{\dagger}\rangle,
\end{align}
\begin{align}\label{eq:Fii}
F_{ii}(t_1,t_2,t_3)=\langle V G^{\dagger}(t_1)\alpha G(t_2)\alpha G(t_3)V^{\dagger}\rangle,
\end{align}
where the retarded Green's function $G(t)=(-i/\hbar)\theta(t)e^{-iHt}$ represents forward time evolution with the free-molecule Hamiltonian $H$ (diagrams $(1,1)a$, $(1,1)b$) and $G^{\dagger}$ represents  backward evolution. $F_i$ involves one forward and two backward evolution periods while $F_{ii}$ contains two forward followed by one backward propagation. $F_i$ and $F_{ii}$ differ by the final state of the matter. In $F_i$ ($F_{ii}$) it is different (the same) from the state after the actinic pulse preparation.

To use entangled light in the IFSRS measurement we first generate frequency and polarization entangled photon pairs via  type-II parametric down conversion (PDC)\cite{Shi03}. The barium borate (BBO) crystal pumped by a femtosecond pulse creates a pair of orthogonally polarized photons which are initially separated by a polarizing beam splitter (BS) in  \ref{fig:setup}d and then directed into two arms of the Hanburry-Brown-Twiss interferometer \cite{HBT56}.  Horizontally polarized beam $s$ interacts with the molecule and serves as a Raman probe in standard FSRS setup, whereas vertically polarized beam $r$ propagates freely and provides a reference. The time-and-frequency resolved detection via ultrafast upconversion  of the photons\cite{Kuz08} in IFSRS provides spectroscopic information about excited state vibrational dynamics of the molecule in the $s$ arm. IFSRS has the following control parameters: the time and frequency parameters of the single photon detectors, which can time the photons with up to $\sim100$ fs resolution\cite{Kuz08}, frequency of the narrowband classical pump pulse $\omega_p$ and the time delay $T$ between the actinic pulse $\mathcal{E}_a$ and  the probe $E_s$.

The photon state produced by PDC has two contributions: a vacuum state and two-photon state with  single photon in the $s$- mode and single photon in the $r$ - mode and is described by the wavefunction
\begin{align}\label{eq:twin}
|\psi\rangle=|0\rangle+\int_{-\infty}^{\infty}d\omega_sd\omega_r\Phi(\omega_s,\omega_r)a_{\omega_s}^{\dagger}a_{\omega_r}^{\dagger}|0\rangle,
\end{align}
where $a_{\omega_s}^{\dagger}$($a_{\omega_r}^{\dagger}$) is the creation operator of a horizontally (vertically) polarized photon and the two-photon amplitude $\Phi(\omega_s,\omega_r)$ is given by \cite{Sch13}
\begin{align}\label{eq:twin1}
&\Phi(\omega_s,\omega_r)=\mathcal{E}_0(\omega_s+\omega_r)\sum_{i\neq j=1}^2\text{sinc}\left(\omega_{s0}T_i/2+\omega_{r0}T_j/2\right)e^{i\omega_{s0}T_i/2+i\omega_{r0}T_j/2},
\end{align}
where $\omega_{k0}=\omega_k-\omega_0$,  $k=s,r$ is the frequency difference between entangled photon and the classical PDC-pump field $\mathcal{E}_0$ that created an entangled pair. In the following simulations we assumed a lorentzian field envelope $\mathcal{E}_0(\omega)=A_0/[\omega-\omega_0+i\sigma_0]$. $T_{j}=(1/v_p-1/v_{j})L$, $k=s,r$ is the time delay acquired by the entangled photon relative to the PDC-pump field due to group velocity dispersion inside the nonlinear crystal. $T_{12}=T_2-T_1$ is  the entanglement time, which controls the timing of the entangled pair. For a narrowband PDC-pump $\mathcal{E}_0(\omega)$ the sum-frequency $\omega_s+\omega_r$ is narrowly distributed around $2\omega_0$ with bandwidth $\sigma_0$. This has been used to selectively prepare double exciton states in two-photon absorption \cite{Lee06,Sch13}. For a broadband PDC-pump the frequency difference $\omega_s-\omega_0$ is narrow with bandwidth $T_j^{-1}$, $j=1,2$. The output state of light in mode $s$ may contain a varying number of photons, depending on the order of the field-matter interaction.

\begin{figure*}[h]
\begin{center}
\includegraphics[trim=0cm 0cm 0cm 0cm,angle=0, width=0.85\textwidth]{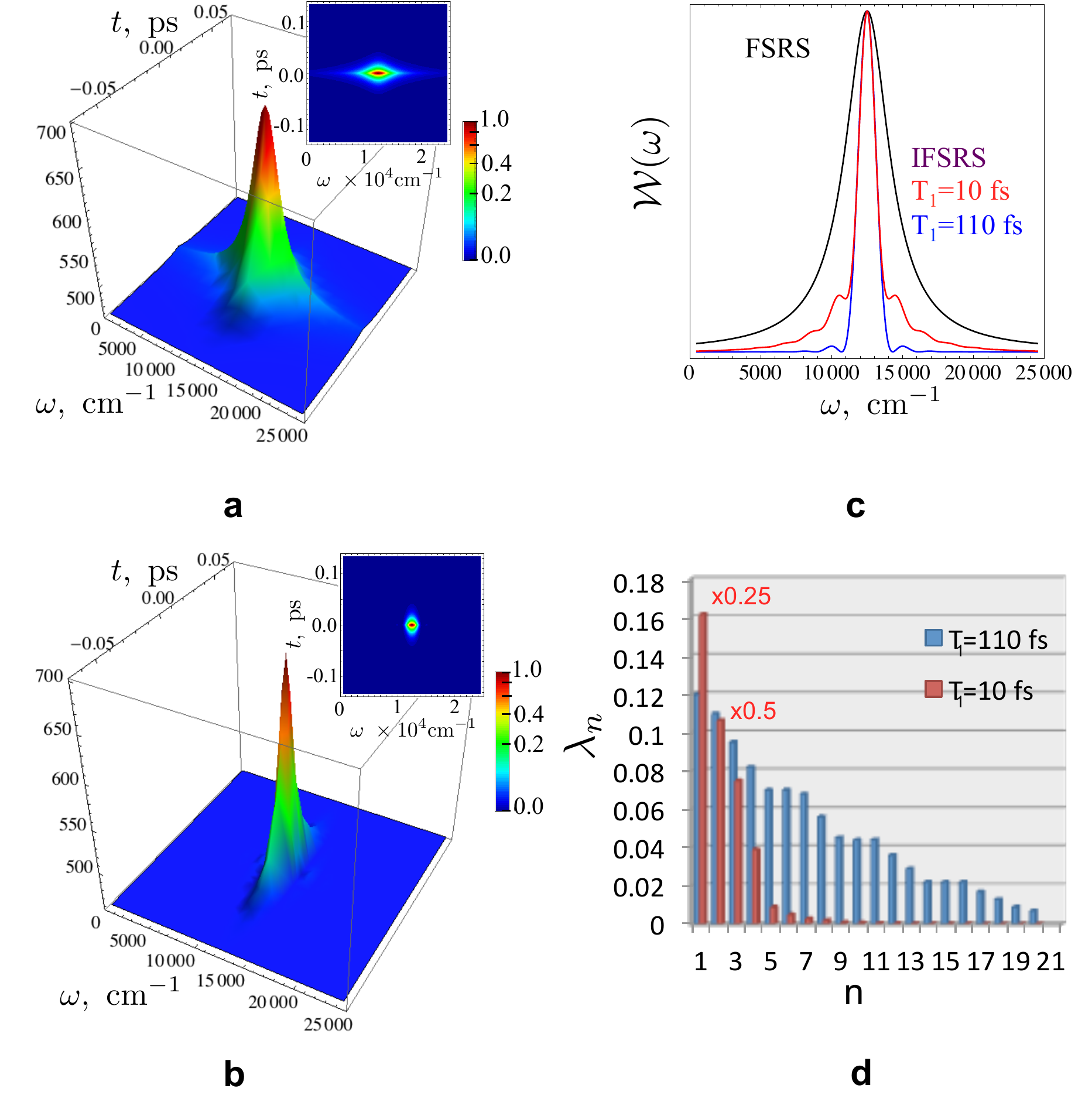}
\end{center}
\caption{(Color online)Left column: $\textbf{a}$ - time-frequency Wigner spectrogram for classical light, $\textbf{b}$ - same as $\textbf{a}$ but for entangled twin state given by Eq. (\ref{eq:twin}). Inserts depict a 2D prejection. Right column: $\textbf{c}$ - window function for FSRS $\mathcal{E}_s^{*}(\omega)\mathcal{E}_s(\omega+i\gamma_a)$ -black, and IFSRS $\Phi^{*}(\omega,\bar{\omega}_r)\Phi(\omega+i\gamma_a,\bar{\omega}_r)$ with $T_1=110$ fs - blue, $T_2=120$ fs, and $T_1=10$ fs, $T_2=120$ fs - red. $\textbf{d}$ - spectrum of the eigenvalues $\lambda_n$ in the Schmidt decomposition (\ref{eq:Sch}) for entangled state with amplitude (\ref{eq:twin1}).The first two eigenvalues $n=1,2$ are scaled with weights $0.25$, $0.5$, respectively. The remaining eigenvalues have no weighted scaling.}
\label{fig:Wig}
\end{figure*}

In general, the twin photon state Eq. (\ref{eq:twin}) is not necessarily entangled. This can determined by the Schmidt decomposition \cite{Law00}
\begin{align}\label{eq:Sch}
\Phi(\omega_s,\omega_r)=\sum_n\sqrt{\lambda_n}\psi_n(\omega_s)\phi_n(\omega_r)
\end{align}
 where $\lambda_k$ are the real positive singular values of $\Phi$ and $\psi_n(\phi_n)$ form an orthonormal set of eigenfunctions of $\int d\omega \Phi(\omega_s,\omega)\Phi^{*}(\omega_r,\omega)$ ($\int d\omega \Phi(\omega,\omega_s)\Phi^{*}(\omega,\omega_r))$ with $\sum_n\lambda_n=1$ for the normalized two-photon state. A separable (unentangled) state has only one non vanishing eigenvalue $\lambda_1=1$ whereas two or more components imply entanglement. The degree of entanglement can be measured by the inverse participation ratio $r_p\equiv(\sum_n\lambda_n^2)^{-1}$. For the two-photon amplitude in Eq. (\ref{eq:twin1}) the rich spectrum of eigenvalues shown in  \ref{fig:Wig}d indicates that the state is highly entangled as $r_p\sim 100$. In addition as can be seen from the insert in  \ref{fig:Wig}a,b, state (\ref{eq:twin1})  is not bound by the Fourier uncertainty $\Delta\omega\Delta t\geq 1$. In the following we study effects of the entanglement on Raman resonances.

\textit{The IFSRS} is given by the rate of a  joint time-and-frequency gated detection of $N_s$ photons in detector $s$ and a single photon in $r$ when both detectors have narrow spectral gating. This is given by 
\begin{align}\label{eq:S10}
&S_{IFSRS}^{(N_s,1)}(\bar{\omega}_{s_1},...,\bar{\omega}_{s_{N_s}},\bar{\omega}_r,\Gamma_i)=\langle\mathcal{T} E_r^{\dagger}(\bar{\omega}_r)E_r(\bar{\omega}_r)\prod_{j=1}^{N_s}E_s^{\dagger}(\bar{\omega}_{s_j})E_s(\bar{\omega}_{s_j})e^{-\frac{i}{\hbar}\int_{-\infty}^{\infty}H'_-(\tau)d\tau}\rangle.
\end{align}
where $\Gamma_i$ represents the incoming light beams, such as central frequency and time, spectral and temporal bandwidth. In the standard Glauber's approach \cite{Glau07} photon counting is calculated in the space of the radiation field using normally ordered field operators. Eq.(\ref{eq:S10}) in contrast operates in the joint matter plus field space and uses time ordered superoperators \cite{Dor141}. This is necessary for the bookkeeping of spectroscopic signals. Both FSRS and IFSRS signals are obtained by the lowest (sixth) order perturbative expansion of Eq. (\ref{eq:S10}) in field-matter interactions (Section \ref{sec:FSRS} of SI) as depicted by the loop diagrams shown in  \ref{fig:setup}c and e, respectively. Measurements with a different number of photons in the $s$ arm are experimentally distinct and are given by different detection windows governed by multipoint correlation function of electric field (red arrows in  \ref{fig:setup}e). Details of the derivations for the field correlation functions for twin entangled state of light are given in Section \ref{sec:field} of SI.

 \ref{fig:Wig} compares field spectrograms which represent the windows created by various fields. \ref{fig:Wig}a depict a time-frequency Wigner function $W_s(\omega,t)=\int_{-\infty}^{\infty}\frac{d\Delta}{2\pi}\mathcal{E}_s^{*}(\omega)\mathcal{E}_s(\omega+\Delta)e^{-i\Delta t}$ for the classical probe field $\mathcal{E}_s$. The time-frequency Fourier uncertainty restricts the frequency resolution for a given time resolution so that $\Delta \omega\Delta t\geq 1$. The Wigner spectrogram $W_q(\omega,t;\bar{\omega}_r)=\int_{-\infty}^{\infty}\frac{d\Delta}{2\pi}\Phi^{*}(\omega,\bar{\omega}_r)\Phi(\omega+\Delta,\bar{\omega}_r)e^{-i\Delta t}$  for the entangled twin photon state is depicted in  \ref{fig:Wig}b. For the same temporal resolution as in FSRS ($\Delta \nu\Delta t\sim 3.7$ ps$\cdot$cm$^{-1}$, which is the Fourier uncertainty for the lorentzian pulses), the spectral resolution of IFSRS is significantly better ($\Delta\nu\Delta t\sim 1.6$ ps$\cdot$cm$^{-1}$). This is possible since the time and frequency resolution for entangled light are not Fourier conjugate variables \cite{Sch13}. The high spectral resolution in the entangled case is governed by $T_j^{-1}$, $j=1,2$ which is narrower than the broadband probe pulse.  \ref{fig:Wig}c demonstrates that the entangled window function $R_q^{(N_s,1)}$ for $N_s=1,2$ (see  Eqs. (\ref{eq:Rq11}), (\ref{eq:Rq21}))  that enters the IFSRS (\ref{eq:Sq3}) yields a much higher spectral resolution than the classical $R_c$ in Eq. (\ref{eq:Rc}).

The molecular information required by the Raman measurements considered here is given by two correlation functions $F_i$ and $F_{ii}$ (see  \ref{fig:setup}c,e and Eqs. (\ref{eq:Fi}) - (\ref{eq:Fii})). These are convoluted with a different detection window for FSRS and IFSRS. $F_i$ and $F_{ii}$ may not be separately detected in FSRS. However in IFSRS the loss  $S_{IFSRS}^{(0,1)}$ and the gain $S_{IFSRS}^{(2,1)}$ Raman signals probe $F_i$ where the final state $c$ can be different from initial state $a$. On the other hand the coincidence counting $S_{IFSRS}^{(1,1)}$ signal is related to $F_{ii}$ (both initial and final states are the same $a$). Interferometric signals can thus separately detect $F_i$ and $F_{ii}$.

\textit{IFSRS for a vibrational mode in a tunneling system.} We had demonstrated  the combined effect of entanglement and interferometric measurement by calculating the signals for the three-level model system undergoing relaxation as depicted in  \ref{fig:setup}a. Once excited by the optical pulse, the vibrational state of the excited electronic state at initial time has frequency $\omega_{a+}=\omega_{a}+\delta$. For a longer time the system tunnels through a barrier at a rate $k$ and assumes a different frequency $\omega_{a-}=\omega_{a}-\delta$. The probability to be in the state with $\omega_{a+}$ decreases exponentially $P_{+}(t)=e^{-kt}$ whereas for $\omega_{a-}$ it grows as $P_{-}(t)=1-e^{-kt}$. This model is mathematically identical to the low temperature limit of Kubo's two-state jump model described by the Stochastic Liouville Equation (SLE) \cite{Kub63, Dor131}. The absorption lineshape is given by 
\begin{align}\label{eq:lin}
S_l(\omega)=-\mathcal{I}\frac{4}{\hbar^2}|\mathcal{E}(\omega)|^2\frac{|\mu_{ag}|^2}{k+2i\delta}\left(\frac{k+i\delta}{\omega-\omega_{a-}+i\gamma_a}+\frac{i\delta}{\omega-\omega_{a+}+i(\gamma_a+k)}\right).
\end{align}
This gives two peaks with combined width governed by dephasing $\gamma_a$ and tunneling rate $k$. Similarly one can derive the corresponding IFSRS signal $S_{IFSRS}^{(N_s,1)}$ with $N_s=0,1,2$ using SLE (see Section \ref{sec:tsj} of SI) which yields

\begin{align}\label{eq:Sq3}
&S_{IFSRS}^{(N_s,1)}(\bar{\omega}_s,\bar{\omega}_r;\omega_p,T)=\mathcal{I}\frac{\mu}{\hbar^4}|\mathcal{E}_p|^2|\mathcal{E}_a|^2\sum_{a,c}\alpha_{ac}^2|\mu_{ag}|^2\notag\\
&\times e^{-2\gamma_aT}\left(R_q^{(N_s,1)}(\bar{\omega}_s,\bar{\omega}_r,2\gamma_a,\bar{\nu}\omega_\nu-i\gamma_a)-\frac{2i\delta e^{-kT}}{k+2i\delta}\right.\notag\\
&\times\left.[R_q^{(N_s,1)}(\bar{\omega}_s,\bar{\omega}_r,2\gamma_a+k,\bar{\nu}\omega_\nu-i\gamma_a)\right.\notag\\
&\left.-R_q^{(N_s,1)}(\bar{\omega}_s,\bar{\omega}_r,2\gamma_a+k,\bar{\nu}\omega_{\bar{\nu}}-i(\gamma_a+k)]\right),
\end{align}
where $\nu=-$ for $N_s=0,2$ and $\nu=+$ for $N_s=1$, $\mu=-$ for $N_s=1,2$ and $\mu=+$ for $N_s=0$. Expressions for the Raman response $R_q^{(N_s,1)}$ which depends on the window created by the quantum field for different photon numbers $N_s$ are given by Eqs. (\ref{eq:Rq01}), (\ref{eq:Rq21}), and (\ref{eq:Rq11}) of SI. The classical FSRS signal (\ref{eq:Sc0}) is given by the similar expression, i.e. $S_{FSRS}^{(c)}=S_{IFSRS}^{(2,1)}[\omega_{\pm}]-S_{IFSRS}^{(2,1)}[-\omega_{\mp}]$ by replacing the entangled detection window $\Phi^{*}(\omega,\bar{\omega}_r)\Phi(\omega+i\gamma,\bar{\omega}_r)$ with a classical one $\mathcal{E}_s^{*}(\omega)\mathcal{E}_s(\omega+i\gamma_a)$.

\begin{figure*}[h]
\begin{center}
\includegraphics[trim=0cm 0cm 0cm 0cm,angle=0, width=0.95\textwidth]{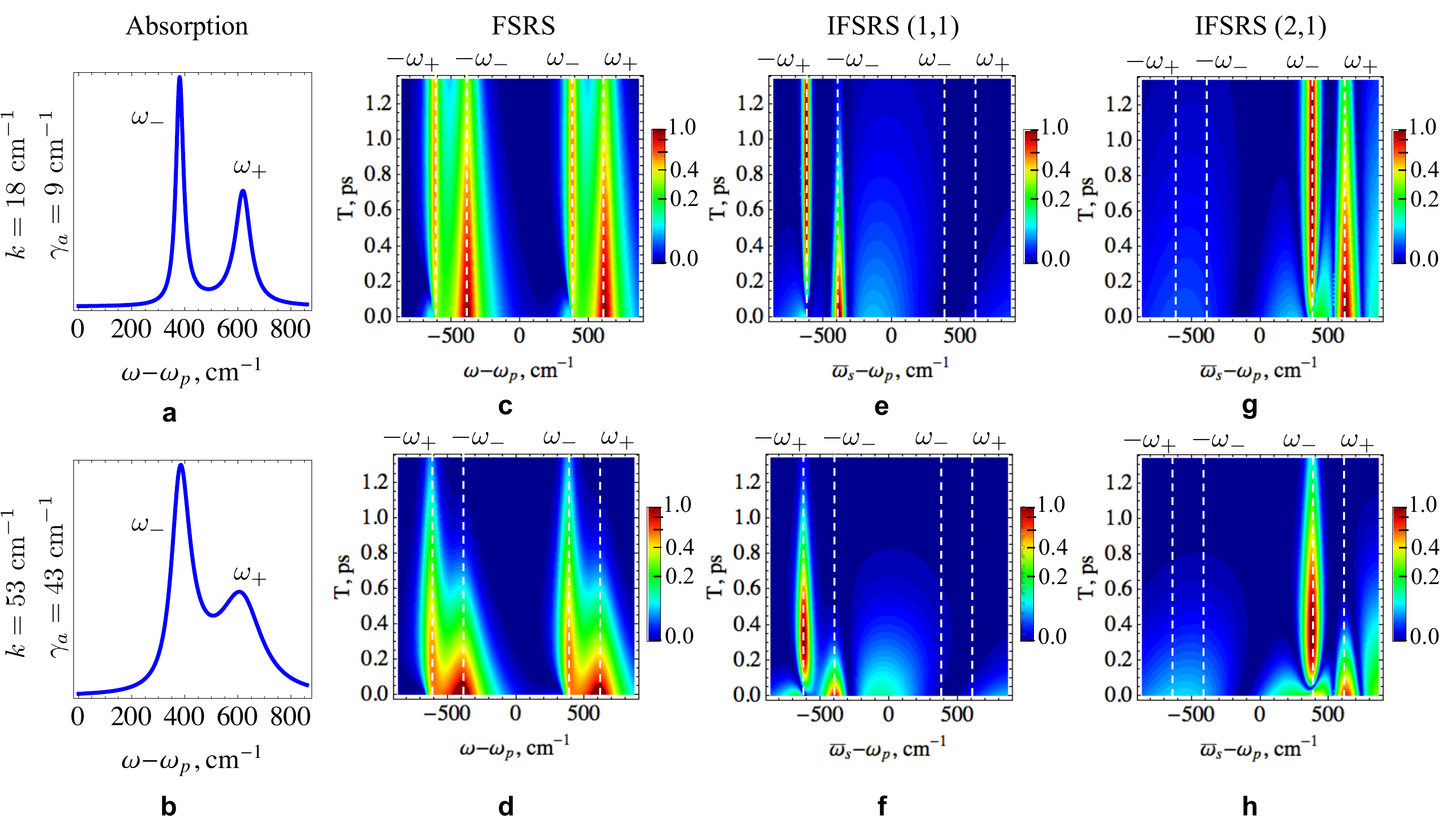}
\end{center}
\caption{(Color online) First column: $\textbf{a}$ - Absorption for a time evolving vibrational mode i vs $\omega-\omega_p$ for slow tunneling rate $k=18$ cm$^{-1}$ and narrow dephasing $\gamma_a=9$ cm$^{-1}$, $\textbf{b}$ - same as $\textbf{a}$ but for fast tunneling rate $k=53$ cm$^{-1}$ and broad dephasing $\gamma_a=43$ cm$^{-1}$. Second column: $\textbf{c}$, $\textbf{d}$ - same as $\textbf{a}$, $\textbf{b}$ but for classical FSRS signal. Third column: $\textbf{e}$ and $\textbf{f}$ - same as $\textbf{a}$, $\textbf{b}$ but for  $S_{IFSRS}^{(1,1)}$, Fourth column: $\textbf{g}$, $\textbf{h}$ - same as $\textbf{a}$, $\textbf{b}$ but for $S_{IFSRS}^{(2,1)}$ given by Eq. (\ref{eq:Sq3}) vs $\bar{\omega}_{s}-\omega_p$. Parameters for the simulations: unperturbed vibration frequency $\omega_{ac}=500$ cm$^{-1}$, level splitting $\delta=120$ cm$^{-1}$, $\omega_p=12500$ cm$^{-1}$, $\bar{\omega}_r=15500$ cm$^{-1}$, and $T_2=120$ fs.  $T_1=10$ fs for $\textbf{c}$, $\textbf{d}$ and $T_1=110 $ fs  for $\textbf{e}$ and $\textbf{f}$. The series of the snapshots (slices of panels $\mathbf{c}$-$\mathbf{h}$) are shown in  \ref{fig:tsjsup} of SI.
}
\label{fig:tsj}
\end{figure*}

 \ref{fig:tsj} compares the classical FSRS signal (Eq. (\ref{eq:Sc3})) with  $S_{IFSRS}^{(1,1)}$ and $S_{IFSRS}^{(2,1)}$ (Eq. (\ref{eq:Sq3})).  For slow modulation and long dephasing time $k,\gamma_a\ll\delta$ the absorption (\ref{fig:tsj}a) has two well-resolved peaks at $\omega_{\pm}$. The classical FSRS shown in  \ref{fig:tsj}c then has one dominant resonance at $\omega_+$ which decays with the delay $T$, whereas the $\omega_-$ peak slowly builds up and dominates at longer $T$. This signal contains both blue- and red-shifted Raman resonances relative to the narrowband pump frequency: $\omega-\omega_p=\pm\omega_{\pm}$. If the modulation  and dephasing rates are comparable to the level splitting $k,\gamma_a\sim\delta$, then the $\omega_{\pm}$ resonances in the absorption (\ref{fig:tsj}b) and the classical FSRS (\ref{fig:tsj}d) become broad and less resolved. It is worth noting, that there is no mirror symmetry between blue- and red- contributions around $\omega=\omega_p$. For $\omega>\omega_p$ the vibration probed by a Raman sequence of pulses at initial time has frequency $\omega-\omega_p=\omega_{+}\equiv\omega_{ac}+\delta$ which gets depopulated with time whereas the transition $\omega-\omega_p=\omega_-=\omega_{ac}-\delta$ gets populated. In the case of $\omega<\omega_p$, the higher vibrational state is given by $-\omega_-$ and the lower vibrational state is $-\omega_+$. So the actual symmetry applies to $\omega_{\pm}\leftrightarrow -\omega_{\mp}$.

We next turn to IFSRS. For slow tunneling  and long dephasing,  $S_{IFSRS}^{(1,1)}$ is similar to the classical FSRS as shown in  \ref{fig:tsj}e. However both temporal and spectral resolutions remain high even when the modulation is fast and the dephasing width is large, as is seen in  \ref{fig:tsj}f. The same applies to the $S_{IFSRS}^{(2,1)}$ signal depicted for slow tunneling -  \ref{fig:tsj}g and fast tunneling -  \ref{fig:tsj}h. Note, that  high resolution for $S_{IFSRS}^{(1,1)}$ and $S_{IFSRS}^{(2,1)}$ signals is achieved for different parameter regimes. At fixed $T_2=120$ fs $S_{IFSRS}^{(1,1)}$ has high resolution at short $T_1=10$ fs whereas  long $T_1=110$ fs works better for $S_{IFSRS}^{(2,1)}$. This difference may be attributed to the selection of field-matter pathways by the different  detection windows of the two signals. Another important difference between the long and short dephasing (top and bottom row in Fig. 3, respectively) is the overall time scale. It follows from Eqs. (\ref{eq:Sc3}) and (\ref{eq:Sq3}) that the signals decay exponentially with the dephasing rate $\sim e^{-2\gamma_aT}$. Therefore for a given range of $0<T<1.3$ ps, the signals with long dephasing (panels c,e, and g in Fig. 3) are stronger than the signals with fast dephasing (panels d,f, and g in Fig. 3).

Apart from the different detection windows, there is another important distinction between IFSRS (Eq. (\ref{eq:Sq3})) and FSRS (\ref{eq:Sc3}) signals. In FSRS, the gain and loss contributions both contain red- and blue- shifted features relative to the narrow pump. The FSRS signal can contain both Stokes and anti Stokes components. Classical FSRS can only distinguish between red and blue contributions. The counting signals, in contrast, can measure separately the gain $S_{IFSRS}^{(2,1)}$ and the loss contributions $S_{IFSRS}^{(0,1)}$ since these are not related to the classical causal response function, which is a specific combination of the quantum matter pathways. Each IFSRS signal is a different combination of pathways that can be expressed uniquely in terms of the left- and right- superoperators.
 
 \textit{Role of entanglement.} We now show that entanglement is essential for the improved resolution of Raman resonances which may not be achieved by classically shaped light. To that end we calculate the IFSRS signals (\ref{eq:Sq3}) for the correlated-separable state of the field\cite{Zhe13} described by the density matrix $\rho_{cor}=\int_{-\infty}^{\infty}d\omega_sd\omega_r|\Phi(\omega_s,\omega_r)|^2|1_{\omega_s},1_{\omega_r}\rangle\langle 1_{\omega_s},1_{\omega_r}|$. This  is the diagonal part of the density matrix corresponding to state (\ref{eq:twin}) with amplitude (\ref{eq:twin1}). This state is not entangled but yields the same single-photon spectrum and shows strong frequency correlations similar to entangled case, and is typically used as a benchmark to quantify entanglement in quantum information processing \cite{Law00}. We further examine the fully separable uncorrelated Fock state given by Eq. (\ref{eq:twin}) with $\Phi_{uncor}(\omega_s,\omega_r)=\Phi_s(\omega_s)\Phi_r(\omega_r)$ with $\Phi_k(\omega_k)=\Phi_0/[\omega_k-\omega_0+i\sigma_0]$, $k=s,r$ with parameters matching the classical probe pulse used in FSRS.

\begin{figure*}[h]
\begin{center}
\includegraphics[trim=0cm 0cm 0cm 0cm,angle=0, width=0.45\textwidth]{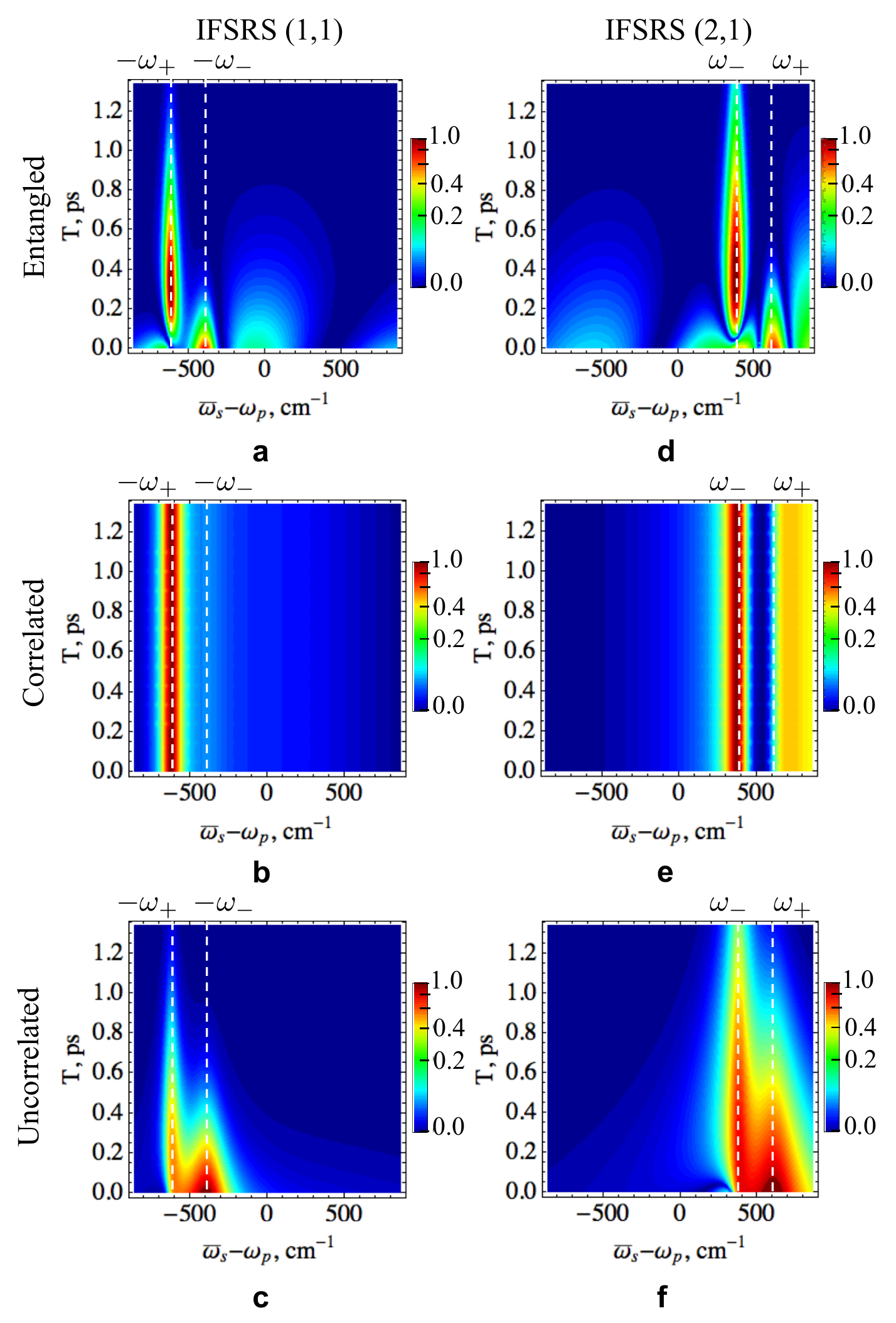}
\end{center}
\caption{(Color online) Left column: $S_{IFSRS}^{(1,1)}$ signal  vs $\bar{\omega}_s-\omega_p$ for entangled state (\ref{eq:twin1}) - $\mathbf{a}$, correlated - $\mathbf{b}$ and uncorrelated - $\mathbf{c}$ separable states. $\mathbf{d}$, $\mathbf{e}$, and $\mathbf{f}$ - same as $\mathbf{a}$, $\mathbf{b}$, and $\mathbf{c}$ but for $S_{IFSRS}^{(2,1)}$ signal. All parameters are the same as in  \ref{fig:tsj}. The corresponding series of the snapshots (slices of these plots) are shown in  \ref{fig:sepsup} of SI.
}
\label{fig:ent}
\end{figure*}

$S_{IFSRS}^{(1,1)}$ for these three states of light are compared in the left column of  \ref{fig:ent}.  \ref{fig:ent}a shows highly resolved Raman resonances for entangled twin state.  The separable correlated state (see \ref{fig:ent}b) has high spectral but no temporal resolution, as expected from a cw time-averaged state in which the photons arrive at any time \cite{Zhe13}.  The separable uncorrelated state (see \ref{fig:ent}c) yields the same resolution as the classical FSRS signal in  \ref{fig:tsj}d since the correlation function of the field factorizes into a product of field amplitudes. Similar results can be obtained for the $S_{IFSRS}^{(2,1)}$ (see  \ref{fig:ent}d, e, and f respectively) Derivations of the IFSRS signals for the correlated and uncorrelated separable states are given in Section \ref{sec:sep} of SI.

In summary, we have demonstrated that stimulated Raman signals with quantum field and interferometric detection better reveal detailed molecular information which is not possible by the standard heterodyne detection of classical fields.

\textbf{Theoretical methods}

In order to use quantum light as a spectroscopic tool for studying complex models of matter, the field-matter interactions must be described in the joint field and matter space. This is done using the superoperator loop diagram formalism \cite{Rah10}. Order by order in field/matter interaction, the signals can be factorized into products of field and matter time-ordered superoperator correlation functions. 

The leading third order signal is governed by a four-point correlation function of the matter. Depending on number of detected photons  this four-point matter correlation function is convoluted with different field correlation function. For $N_s=0$, $N_r=1$  Eq. (\ref{eq:S10}) is given by four-point correlation function for a quantum field.  For a twin photon state it can be factorized as $\langle\psi|E_s^{\dagger}(\omega_a)E_r^{\dagger}(\omega_b)E_r(\omega_c)E_s(\omega_d)|\psi\rangle=\Phi^{*}(\omega_a,\omega_b)\Phi(\omega_c,\omega_d)$. For $N_s=2$ it is given by eight-point (see Eq. (\ref{eq:S210})), whereas for $N_s=1$ it is governed by a six-point field correlation function as shown in Eqs. (\ref{eq:S11a0}) - (\ref{eq:S11b0}). 
For the two-photon state, normally-ordered field correlation functions with more than 4 fields vanish since extra annihilation operators act on the vacuum state. Therefore, the higher order non-normally ordered field correlation functions can be recast as a four-point correlation function times multiple field commutators, which are given by $[E_s(\omega),E_s^{\dagger}(\omega')]=\mathcal{D}(\omega_p)\delta(\omega-\omega')$ where $\mathcal{D}(\omega_s)\simeq\mathcal{D}(\omega_p)$ is the constant which is assumed to be a flat function of its argument for a normalized two photon state. Therefore for $N_s=2$ and $N_s=1$ the signal is proportional to $\mathcal{D}^2(\omega_p)$ and $\mathcal{D}(\omega_p)$, respectively.  All three IFSRS signals with $N_s=0,1,2$ scale as $\sim|\mathcal{E}_0|^2|\mathcal{E}_a|^2|\mathcal{E}_p|^2$ with field intensity, the same as classical FSRS even though a different number of fields contribute to the detection.

\textbf{Acknowledgements}

We gratefully acknowledge the support of the National Science Foundation through Grant No. CHE-1058791 and computations are supported by CHE-0840513,  the Chemical Sciences, Geosciences and Biosciences Division, Office of Basic Energy Sciences, Office of Science and US Department of Energy, National Institute of Health Grant No. GM-59230. F.S. would like to thank the German National Academic Foundation.


\clearpage

\clearpage

\begin{center}
{\large {\ }\textbf{SUPPLEMENTARY INFORMATION:\\
Stimulated Raman Spectroscopy with Entangled Light; Enhanced Resolution and Pathway Selection}}
\end{center}





\vspace{0.5cm}





\section{Time-and-frequency resolved photon counting signals}\label{sec:tf}

The joint time-and-frequency gated rate of detecting $N_s$ photons in detector $s$ and $N_r$ in $r$ is formally given by\cite{Dor12}
\begin{align}\label{eq:S0}
S_{IS}^{(N_s,N_r)}(\Gamma_{s_1},...,\Gamma_{s_{N_s}},&\Gamma_{r_1},...,\Gamma_{r_{N_r}};\Gamma_i)=\int_{-\infty}^{\infty}dt_{s_1}...\int_{-\infty}^{\infty}dt_{s_{N_s}}\int_{-\infty}^{\infty}dt_{r_1}...\int_{-\infty}^{\infty}dt_{r_{N_r}}\notag\\
&\times\langle \mathcal{T}\prod_{k=1}^{N_r}E_{rR}^{(tf)\dagger}(t_{r_k})E_{rL}^{(tf)}(t_{r_k})\prod_{j=1}^{N_s}E_{sR}^{(tf)\dagger}(t_{s_j})E_{sL}^{(tf)}(t_{s_j})e^{-\frac{i}{\hbar}\int_{-\infty}^{\infty}H'_-(\tau)d\tau}\rangle,
\end{align}
where $\Gamma_{s_j}$, $\Gamma_{r_k}$, $j=1,...,N_s$, $k=1,...,N_r$ denote the set of parameters that characterize the various photons detected in $s$ and $r$, and $\Gamma_i$ represents the incoming light beams. In the following we only consider measurements with a single reference photon $N_r=1$. In Eq. (\ref{eq:S0}) $\langle...\rangle=\text{tr}[...\rho]$ with $\rho$ being the density operator of the entire system, and $H_-'$ is the Hamiltonian superoperator in the interaction picture. Superoperators provide a convenient bookkeeping of time-ordered Green's functions. With every ordinary operator $A$ we associate two superoperators \cite{Har08} defined by their action on an ordinary operator $X$ as $A_L=AX$  acting from left, $A_R=XA$ (right). We further define the symmetric and antisymmetric combinations $A_+=\frac{1}{2}(A_L+A_R)$, $A_-=A_L-A_R$, $\mathcal{T}$ denotes superoperator time ordering.

The coincidence signal Eq. (\ref{eq:S0}) depends on the time-and-frequency gated electric field $E^{(tf)}$. The detector with input located at $r_D$ is represented by a time gate $F_t$ centered at $\bar{t}$ followed by a frequency gate $F_f$ centered at $\bar{\omega}$ \cite{Sto94}. The gated field can be written as
\begin{align}\label{eq:eft}
&E_j^{(tf)}(\bar{t},\bar{\omega};t)=\int_{-\infty}^{\infty}dt'F_f^{(j)}(t-t',\bar{\omega})F_t^{(j)}(t',\bar{t})E_j(t').
\end{align}
The parameters of the time $F_t$ and frequency $F_f$  gates can be varied independently. However the combined temporal and spectral resolution for the signal (\ref{eq:S0}) always satisfies Fourier uncertainty $\Delta \omega\Delta t\geq 1$ \cite{Dor141}.

Hereafter we consider two types of signals. The first has a sharp time gate $F_t^{(s_j)}(\bar{t}_{s_j}t_{s_j}')=\delta(t_s'-\bar{t}_{s_j})$,  $j=1,...,N_s$ for the $s$  detector and narrow frequency gate for the $r$ detector $F_f^{(r)}(\bar{\omega}_r;\omega_r')=\delta(\omega_r'-\bar{\omega}_r)$. The signal (\ref{eq:S0}) then reads
\begin{align}\label{eq:S1t}
S_{IS}^{(N_s,1)}(\bar{t}_{s_1},...,\bar{t}_{s_{N_s}},\bar{\omega}_r,\Gamma_i)=\langle\mathcal{T} E_r^{\dagger}(\bar{\omega}_r)E_r(\bar{\omega}_r)\prod_{j=1}^{N_s}E_s^{\dagger}(\bar{t}_{s_j})E_s(\bar{t}_{s_j})e^{-\frac{i}{\hbar}\int_{-\infty}^{\infty}H'_-(\tau)d\tau}\rangle,
\end{align}
where $\bar{t}_{s_1}<\bar{t}_{s_2}<...<\bar{t}_{s_{N_s}}$. 

The second type of signal is obtained by replacing the time gate for detector $s$ by the frequency gate   $F_f^{(s_j)}(\bar{\omega}_{s_j};\omega_{s_j}')=\delta(\omega_s'-\bar{\omega}_{s_j})$,   $j=1,...,N_s$, retaining the same narrow frequency gate for detector $r$. Eq. (\ref{eq:S10}) then follows from (\ref{eq:S0}).

\section{Femtosecond Stimulated Raman Signals}\label{sec:FSRS}

\subsection{FSRS}

The stimulated FSRS signal obtained in the setup shown in Fig. \ref{fig:setup}b with frequency dispersed detection of the probe is given by
\begin{align}\label{eq:Sc0}
S_{FSRS}(\omega,T)=\frac{2}{\hbar}\mathcal{I}\int_{-\infty}^{\infty}dte^{i\omega(t-T)}\langle\mathcal{E}_s^{*}(\omega)\mathcal{E}_p(t)\alpha(t)e^{-\frac{i}{\hbar}\int H'_-(\tau)d\tau}\rangle,
\end{align}
where $\mathcal{I}$ denotes the imaginary part and $\mathcal{E}_s=\langle E_s\rangle$ is expectation value of the probe field operator with respect to classical state of light. We next expand the signal (\ref{eq:Sc0}) to to sixth order $\sim \mathcal{E}_s^2\mathcal{E}_p^2\mathcal{E}_a^2$. Thus, the classical FSRS signal illustrated by the two diagrams shown in Fig. \ref{fig:setup}c is given by $S_{FSRS}^{(c)}(\omega,T)=S_{FSRS}^{(i)}(\omega,T)+S_{FSRS}^{(ii)}(\omega,T)$ where
\begin{align}\label{eq:Sisr1}
S_{FSRS}^{(i)}(\omega,T)&=\frac{2}{\hbar}\mathcal{I}\int_{-\infty}^{\infty}dte^{i\omega(t-T)}\int_{-\infty}^td\tau_1\int_{-\infty}^tdt'\int_{-\infty}^{t'}d\tau_2 \mathcal{E}_p(t)\mathcal{E}_p^{*}(t')\mathcal{E}_a^{*}(\tau_2)\mathcal{E}_a(\tau_1)\notag\\
&\times\mathcal{E}_s^{*}(\omega)\mathcal{E}_s(t') F_i(t'-\tau_2,t-t',t-\tau_1),
\end{align}
\begin{align}\label{eq:Siisr1}
S_{FSRS}^{(ii)}(\omega,T)&=\frac{2}{\hbar}\mathcal{I}\int_{-\infty}^{\infty}dte^{i\omega(t-T)}\int_{-\infty}^td\tau_2\int_{-\infty}^tdt'\int_{-\infty}^{t'}d\tau_1\mathcal{E}_p(t)\mathcal{E}_p^{*}(t')\mathcal{E}_a(\tau_1)\mathcal{E}_a^{*}(\tau_2)\notag\\
&\times \mathcal{E}_s^{*}(\omega)\mathcal{E}_s(t')F_{ii}(t-\tau_2,t-t',t'-\tau_1).
\end{align}

\subsection{IFSRS}

Expansion of Eq. (\ref{eq:S10}) for the field matter interactions depicted by loop diagrams in Fig. \ref{fig:setup}e yields for $N_s=0$ - Raman loss (no photon in the molecular arm)
\begin{align}\label{eq:S010}
S_{IFSRS}^{(0,1)}(\bar{\omega}_r;T)&=\mathcal{I}\frac{1}{\hbar}\int_{-\infty}^{\infty}dt\int_{-\infty}^{\infty}dt'\int_{-\infty}^{t}d\tau_1\int_{-\infty}^{t'}d\tau_2\mathcal{E}_p(t')\mathcal{E}_p^{*}(t)\mathcal{E}_a(\tau_1)\mathcal{E}_a^{*}(\tau_2)\notag\\
&\times\langle \mathcal{T}E_s^{\dagger}(t')\tilde{E}_r^{\dagger}(\bar{\omega}_r)\tilde{E}_r(\bar{\omega}_r)E_s(t)\rangle F_i(t'-\tau_2,t-t',t-\tau_1).
\end{align}
To make sure that there is no photon at detector $s$ we had integrated over its full bandwidth, thus eliminating the dependence on detector parameters.  For the Raman gain $N_s=2$ signal we get
\begin{align}\label{eq:S210}
&S_{IFSRS}^{(2,1)}(\bar{t}_{s_1},\bar{t}_{s_2},\bar{\omega}_r;T)=\mathcal{I}\frac{1}{\hbar}\int_{-\infty}^{\bar{t}_{s_1}}dt\int_{-\infty}^{\bar{t}_{s_1}}dt'\int_{-\infty}^{t}d\tau_1\int_{-\infty}^{t'}d\tau_2\mathcal{E}_p(t)\mathcal{E}_p^{*}(t')\mathcal{E}_a(\tau_1)\mathcal{E}_a^{*}(\tau_2)\notag\\
&\times\langle \mathcal{T}E_s(t')\tilde{E}_{s}^{\dagger}(\bar{t}_{s_1})\tilde{E}_{s}^{\dagger}(\bar{t}_{s_2})\tilde{E}_r^{\dagger}(\bar{\omega}_r)\tilde{E}_r(\bar{\omega}_r)\tilde{E}_{s}(\bar{t}_{s_2})\tilde{E}_{s}(\bar{t}_{s_1})E_s^{\dagger}(t)\rangle F_i(t'-\tau_2,t-t',t-\tau_1).
\end{align}
Similarly for $N_s=1$ we obtain
\begin{align}\label{eq:S11a0}
S_{IFSRS}^{(1,1)a}(\bar{t}_s,\bar{\omega}_r;T)&=-\mathcal{I}\frac{1}{\hbar}\int_{-\infty}^{\bar{t}_s}dt\int_{-\infty}^tdt'\int_{-\infty}^{t'}d\tau_1\int_{-\infty}^{\bar{t}_s}d\tau_2\mathcal{E}_p(t)\mathcal{E}_p^{*}(t')\mathcal{E}_a(\tau_1)\mathcal{E}_a^{*}(\tau_2)\notag\\
&\times\langle \mathcal{T}\tilde{E}_s^{\dagger}(\bar{t}_s)\tilde{E}_r^{\dagger}(\bar{\omega}_r)\tilde{E}_r(\bar{\omega}_r)\tilde{E}_s(\bar{t}_s)E_s^{\dagger}(t)E_s(t')\rangle F_{ii}(t-\tau_2,t-t',t'-\tau_1),
\end{align}
\begin{align}\label{eq:S11b0}
S_{IFSRS}^{(1,1)b}(\bar{t}_s,\bar{\omega}_r;T)&=-\mathcal{I}\frac{1}{\hbar}\int_{-\infty}^{\bar{t}_s}dt\int_{-\infty}^tdt'\int_{-\infty}^{t'}d\tau_1\int_{-\infty}^{\bar{t}_s}d\tau_2\mathcal{E}_p(t')\mathcal{E}_p^{*}(t)\mathcal{E}_a(\tau_1)\mathcal{E}_a^{*}(\tau_2)\notag\\
&\times\langle \mathcal{T}\tilde{E}_s^{\dagger}(\bar{t}_s)\tilde{E}_r^{\dagger}(\bar{\omega}_r)\tilde{E}_r(\bar{\omega}_r)\tilde{E}_s(\bar{t}_s)E_s(t)E_s^{\dagger}(t')\rangle F_{ii}(t-\tau_2,t-t',t'-\tau_1),
\end{align}
where $\tilde{E}=E^{tf}$ is the gated field and $\mathcal{R}$ denotes the real part that comes from the complex conjugate diagrams shown in Fig. \ref{fig:setup}e.

One can similarly express the signal (\ref{eq:S10})  with $N_s=2$ when both $s$ and $r$ detectors have narrow frequency gates 
\begin{align}\label{eq:S2100}
&S_{IFSRS}^{(2,1)}(\bar{\omega}_{s_1},\bar{\omega}_{s_2},\bar{\omega}_r;T)=\mathcal{I}\frac{1}{\hbar}\int_{-\infty}^{\infty}d\bar{t}_{s_1}e^{i\bar{\omega}_{s_1}(\bar{t}_{s_1}-T)}\int_{-\infty}^{\bar{t}_{s_1}}dt\int_{-\infty}^{t}dt'\int_{-\infty}^{t}d\tau_1\int_{-\infty}^{t'}d\tau_2\mathcal{E}_p(t)\mathcal{E}_p^{*}(t')\mathcal{E}_a(\tau_1)\mathcal{E}_a^{*}(\tau_2)\notag\\
&\times\langle \mathcal{T}E_s(t')\tilde{E}_{s}^{\dagger}(\bar{\omega}_{s_1})\tilde{E}_{s}^{\dagger}(\bar{\omega}_{s_2})\tilde{E}_r^{\dagger}(\bar{\omega}_r)\tilde{E}_r(\bar{\omega}_r)\tilde{E}_{s}(\bar{\omega}_{s_2})\tilde{E}_{s}(\bar{t}_{s_1})E_s^{\dagger}(t)\rangle F_i(t'-\tau_2,t-t',t-\tau_1).
\end{align}

Similarly the $N_s=1$ signal  is given by
\begin{align}\label{eq:S11a1}
S_{IFSRS}^{(1,1)a}(\bar{\omega}_s,\bar{\omega}_r;T)&=-\mathcal{I}\frac{1}{\hbar}\int_{-\infty}^{\infty}dt_s'e^{i\bar{\omega}_s(t_s'-T)}\int_{-\infty}^{t_s'}dt\int_{-\infty}^tdt'\int_{-\infty}^{t'}d\tau_1\int_{-\infty}^{t_s'}d\tau_2\mathcal{E}_p(t)\mathcal{E}_p^{*}(t')\mathcal{E}_a(\tau_1)\mathcal{E}_a^{*}(\tau_2)\notag\\
&\times\langle \mathcal{T}\tilde{E}_s^{\dagger}(\bar{\omega}_s)\tilde{E}_r^{\dagger}(\bar{\omega}_r)\tilde{E}_r(\bar{\omega}_r)\tilde{E}_s(t_s')E_s^{\dagger}(t)E_s(t')\rangle F_{ii}(t-\tau_2,t-t',t'-\tau_1),
\end{align}
\begin{align}\label{eq:S11b1}
S_{IFSRS}^{(1,1)b}(\bar{\omega}_s,\bar{\omega}_r;T)&=-\mathcal{I}\frac{1}{\hbar}\int_{-\infty}^{\infty}dt_s'e^{i\bar{\omega}_s(t_s'-T)}\int_{-\infty}^{t_s'}dt\int_{-\infty}^tdt'\int_{-\infty}^{t'}d\tau_1\int_{-\infty}^{t_s'}d\tau_2\mathcal{E}_p(t')\mathcal{E}_p^{*}(t)\mathcal{E}_a(\tau_1)\mathcal{E}_a^{*}(\tau_2)\notag\\
&\times\langle \mathcal{T}\tilde{E}_s^{\dagger}(\bar{\omega}_s)\tilde{E}_r^{\dagger}(\bar{\omega}_r)\tilde{E}_r(\bar{\omega}_r)\tilde{E}_s(t_s')E_s(t)E_s^{\dagger}(t')\rangle F_{ii}(t-\tau_2,t-t',t'-\tau_1).
\end{align}

\section{Field correlation functions of entangled light for various IFSRS signals}\label{sec:field}

All our signals (\ref{eq:S010}) - (\ref{eq:S11b1}) involve products of multiple fields and four point correlation functions of matter. For $N_s=0$, $N_r=1$ the four-point correlation function in Eq. (\ref{eq:S010}) for a quantum field  in a twin photon state can be factorized as
\begin{align}\label{eq:4pt}
\langle\psi|E_s^{\dagger}(t')\tilde{E}_r^{\dagger}(\bar{\omega}_r)\tilde{E}_r(\bar{\omega}_r)E_s(t)|\psi\rangle&=\langle\psi| E_s^{\dagger}(t')\tilde{E}_r^{\dagger}(\bar{\omega}_r)|0\rangle\langle 0|\tilde{E}_r(\bar{\omega}_r)E_s(t)|\psi\rangle.
\end{align}

 The twin state of light is described by the wavefunction (\ref{eq:twin}). The two point correlation functions in Eq. (\ref{eq:4pt}) are then given by
\begin{align}\label{eq:EE}
\langle 0| \tilde{E}_r(\bar{\omega}_r)E_s(t)|\psi\rangle=\Phi(t,\bar{\omega}_r)=\int_{-\infty}^{\infty}\frac{d\omega}{2\pi}e^{-i\omega t}\Phi(\omega,\bar{\omega}_r),
\end{align}
and
\begin{align}\label{eq:E1E1}
\langle\psi|E_s^{\dagger}(t')\tilde{E}_r^{\dagger}(\bar{\omega}_r)|0\rangle=\Phi^{*}(t',\bar{\omega}_r)=\int_{-\infty}^{\infty}\frac{d\omega'}{2\pi}e^{i\omega' t'}\Phi^{*}(\omega',\bar{\omega}_r),.
\end{align}
The four point correlation function in Eq. (\ref{eq:4pt}) is finally given by
\begin{align}\label{eq:4pt1}
\langle\psi|E_s^{\dagger}(t')\tilde{E}_r^{\dagger}(\bar{\omega}_r)\tilde{E}_r(\bar{\omega}_r)E_s(t)|\psi\rangle=\Phi^{*}(t',\bar{\omega}_r)\Phi(t,\bar{\omega}_r).
\end{align}

We next turn to the $N_s=2$, $N_r=1$ signal (Eq. (\ref{eq:S210})). In order to evaluate the necessary eight-point field correlation function one can recast it in a  normally ordered form (all annihilation operators are to the right of the creation operators).
For a twin photon state normally ordered correlation functions of the field with more than 4 fields vanish since extra annihilation operators act on the vacuum. Therefore, the eight-point field correlation function in (\ref{eq:S210}) can be recast as a four-point correlation function similar to Eq. (\ref{eq:4pt1}) times two field commutators
\begin{align}
[E_s(t),E_s^{\dagger}(t')]=\mathcal{D}(\omega_p)\delta(t-t'),\quad [E_s(\omega),E_s^{\dagger}(\omega')]=\mathcal{D}(\omega_p)\delta(\omega-\omega'),
\end{align}
where $\mathcal{D}(\omega_s)\simeq\mathcal{D}(\omega_p\pm\omega_{ac})\simeq\mathcal{D}(\omega_p)$ is the normalization constant which is assumed to be a flat function of its argument.
We thus obtain
\begin{align}
&\langle \psi|E_s(t')\tilde{E}_{s}^{\dagger}(\bar{t}_{s_1})\tilde{E}_{s}^{\dagger}(\bar{t}_{s_2})\tilde{E}_r^{\dagger}(\bar{\omega}_r)\tilde{E}_r(\bar{\omega}_r)\tilde{E}_{s}(\bar{t}_{s_2})\tilde{E}_{s}(\bar{t}_{s_1})E_s^{\dagger}(t)|\psi\rangle\notag\\
&=\mathcal{D}^2(\omega_p)[\Phi^{*}(\bar{t}_{s_1},\bar{\omega}_r)\Phi(\bar{t}_{s_1},\bar{\omega}_r)\delta(t-\bar{t}_{s_2})\delta(t'-\bar{t}_{s_2})\notag\\
&+\Phi^{*}(\bar{t}_{s_2},\bar{\omega}_r)\Phi(\bar{t}_{s_2},\bar{\omega}_r)\delta(t-\bar{t}_{s_1})\delta(t'-\bar{t}_{s_1})\notag\\
&+\Phi^{*}(\bar{t}_{s_1},\bar{\omega}_r)\Phi(\bar{t}_{s_2},\bar{\omega}_r)\delta(t-\bar{t}_{s_1})\delta(t'-\bar{t}_{s_2})\notag\\
&+\Phi^{*}(\bar{t}_{s_2},\bar{\omega}_r)\Phi(\bar{t}_{s_1},\bar{\omega}_r)\delta(t-\bar{t}_{s_2})\delta(t'-\bar{t}_{s_1})].
\end{align}

Finally we turn to the $N_s=N_r=1$ signal (Eq. (\ref{eq:S11a0})) which is governed by a six-point correlation function. The only contribution to this correlation function comes from the four-point correlation function of normally ordered fields multiplied by a commutator of the field:
\begin{align}\label{eq:6pt1a}
&\langle\tilde{E}_s^{\dagger}(\bar{t}_s)\tilde{E}_r^{\dagger}(\bar{\omega}_r)\tilde{E}_r(\bar{\omega}_r)\tilde{E}_s(\bar{t}_s)E_s^{\dagger}(t)E_s(t')\rangle=\mathcal{D}(\omega_p)\Phi^{*}(\bar{t}_s,\bar{\omega}_r)\Phi(t',\bar{\omega}_r)\delta(t-\bar{t}_s).
\end{align}

Similarly for Eq. (\ref{eq:S11b0}) we obtain
\begin{align}\label{eq:6pt1b}
\langle\tilde{E}_s^{\dagger}(\bar{t}_s)\tilde{E}_r^{\dagger}(\bar{\omega}_r)\tilde{E}_r(\bar{\omega}_r)\tilde{E}_s(\bar{t}_s)E_s(t)E_s^{\dagger}(t')\rangle&=\mathcal{D}(\omega_p)\Phi^{*}(\bar{t}_s,\bar{\omega}_r)\Phi(\bar{t}_s,\bar{\omega}_r)\delta(t-t')\notag\\
&+\mathcal{D}(\omega_p)\Phi^{*}(\bar{t}_s,\bar{\omega}_r)\Phi(t,\bar{\omega}_r)\delta(t'-\bar{t}_s).
\end{align}
It is worth noting that according to diagram $(1,1)b$ in Fig. \ref{fig:setup}e, the  signal (\ref{eq:S11b0}) has $\bar{t}_s>t>t'$. However it follows from the  second term in Eq. (\ref{eq:6pt1b}) that $t'=\bar{t}_s$, i.e. $t<t'$. Therefore this term does not contribute to the signal.
Note also that like the classical FSRS signal all three IFSRS signals (\ref{eq:4pt1}) - (\ref{eq:6pt1b}) scale linearly with the classical pump intensity $S_{IFSRS}\propto |A_0|^2$, even though they are governed by different number of fields contribution to the detection events. 

\section{Stimulated Raman signals for a vibrational tunneling model}\label{sec:tsj}


Below we discuss a  dynamical vibration model. We consider a single vibrational mode that can assume two states with  frequencies $\omega_{\pm}=\omega_{ac}\pm\delta$ and a tunneling rate between them $k$. Following the stochastic Liouville equation approach outlined in \cite{Dor131}, the absorption lineshape for such a system is given by Eq. (\ref{eq:lin}).

The matter correlation functions in Eqs. (\ref{eq:S010}) - (\ref{eq:S11b1}) $F_j(t_1,t_2,t_3)\to\mathcal{F}_j(t_1-t_2,t_2-t_3)$, $j=i$, $ii$ where
\begin{align}\label{eq:F12l}
& \mathcal{F}_i(t_1,t_2)=\frac{i}{\hbar^3}\sum_{a,c}|\mu_{ag}|^2\alpha_{ac}^2\theta(t_1)\theta(t_2)e^{-\gamma_a(t_1+2t_2)}\left[e^{-i\omega_-t_1}-\frac{2i\delta}{k+2i\delta}e^{-kt_2}\left(e^{-i\omega_-t_1}-e^{-(k+i\omega_+)t_1}\right)\right],
 \end{align}
 \begin{align}\label{eq:F122l}
& \mathcal{F}_{ii}(t_1,t_2)=-\frac{i}{\hbar^3}\sum_{a,c}|\mu_{ag}|^2\alpha_{ac}^2\theta(t_1)\theta(t_2)e^{-\gamma_a(t_1+2t_2)}\left[e^{i\omega_+t_1}-\frac{2i\delta}{k+2i\delta}e^{-kt_2}\left(e^{i\omega_+t_1}-e^{-(k-i\omega_-)t_1}\right)\right].
 \end{align}

Using (\ref{eq:F12l}) the classical FSRS signal (\ref{eq:Sisr1}) - (\ref{eq:Siisr1}) reads
\begin{align}\label{eq:Sc3}
&S_{FSRS}^{(c)}(\omega,T)=-\mathcal{I}\frac{2}{\hbar^4}|\mathcal{E}_p|^2|\mathcal{E}_a|^2\sum_{a,c}\alpha_{ac}^2|\mu_{ag}|^2e^{-2\gamma_aT}\left[R_c(\omega,2\gamma_a,\omega_--i\gamma_a)\right.\notag\\
&\left.-\frac{2i\delta e^{-kT}}{k+2i\delta}[R_c(\omega,2\gamma_a+k,\omega_--i\gamma_a)-R_c(\omega,2\gamma_a+k,\omega_+-i(\gamma_a+k)]-(\omega_{\pm}\leftrightarrow-\omega_{\mp})\right],
\end{align}
 where
\begin{align}\label{eq:Rc}
R_c(\omega,\gamma,\Omega)=\frac{\mathcal{E}_s^{*}(\omega)\mathcal{E}_s(\omega+i\gamma)}{\omega-\omega_p-\Omega}
\end{align}
is the Raman response gated by the classical field.

The IFSRS  $S_{IFSRS}^{(0,1)}$ signal (\ref{eq:S010}) reads
\begin{align}\label{eq:TSJ01}
&S_{IFSRS}^{(0,1)}(\bar{\omega}_r;\omega_p,T)=\mathcal{I}\frac{1}{\hbar^4}|\mathcal{E}_p|^2|\mathcal{E}_a|^2\sum_{a,c}\alpha_{ac}^2|\mu_{ag}|^2e^{-2\gamma_aT}\left(R_q^{(0,1)}(\bar{\omega}_s,\bar{\omega}_r,2\gamma_a,\omega_--i\gamma_a)\right.\notag\\
&\left.-\frac{2i\delta e^{-kT}}{k+2i\delta}[R_q^{(0,1)}(\bar{\omega}_s,\bar{\omega}_r,2\gamma_a+k,\omega_--i\gamma_a)-R_q^{(0,1)}(\bar{\omega}_s,\bar{\omega}_r,2\gamma_a+k,\omega_+-i(\gamma_a+k)]\right),
\end{align}
where
\begin{align}\label{eq:Rq01}
R_q^{(0,1)}(\bar{\omega}_s,\bar{\omega}_r,\gamma,\Omega)=\int_{-\infty}^{\infty}\frac{d\omega}{2\pi}\frac{\Phi^{*}(\omega,\bar{\omega}_r)\Phi(\omega+i\gamma,\bar{\omega}_r)}{\omega-\omega_p-\Omega}
\end{align}
is a Raman response gate by the quantum entangled field.

We next turn to the time-gated $S_{IFSRS}^{(2,1)}$ signal (\ref{eq:S210})
\begin{align}
&S_{IFSRS}^{(2,1)}(\bar{t}_{s_1},\bar{t}_{s_2},\bar{\omega}_r;\omega_p,T)=\mathcal{I}\frac{1}{\hbar}\mathcal{D}^2(\omega_p)|\mathcal{E}_a|^2|\mathcal{E}_p|^2\notag\\
&\times\sum_{i,j=1,2}\Phi^{*}(\bar{t}_{s_i},\bar{\omega}_r)\Phi(\bar{t}_{s_j},\bar{\omega}_r)e^{-i\omega_p(t_{s_j}-t_{s_i})}\mathcal{F}_i(t_{s_j}-t_{s_i},t_{s_i}).
\end{align}
The corresponding frequency-gated $S_{IFSRS}^{(2,1)}$ signal (\ref{eq:S2100}) is given by
\begin{align}\label{eq:TSJ21}
&S_{IFSRS}^{(2,1)}(\bar{\omega}_{s_1},\bar{\omega}_{s_2},\bar{\omega}_r;\omega_p,T)=\mathcal{I}\frac{1}{\hbar^4}\mathcal{D}^2(\omega_p)|\mathcal{E}_a|^2|\mathcal{E}_p|^2\sum_{a,c}|\mu_{ag}|^2\alpha_{ac}^2\Phi^{*}(\bar{\omega}_{s_1},\bar{\omega}_r)\notag\\
&\times\left(\Phi(\bar{\omega}_{s_1},\bar{\omega}_r)\left[\frac{1}{2\gamma_a}\frac{1}{\bar{\omega}_{s_2}-\omega_p-\omega_-+i\gamma_a}-\frac{2i\delta}{k+2i\delta}\frac{1}{k+2\gamma_a}\right.\right.\notag\\
&\left.\times\left.\left(\frac{1}{\bar{\omega}_{s_2}-\omega_p-\omega_-+i\gamma_a}-\frac{1}{\bar{\omega}_{s_2}-\omega_p-\omega_++i(\gamma_a+k)}\right)\right]+i\Phi(\bar{\omega}_{s_2},\bar{\omega}_r)e^{i(\bar{\omega}_{s_2}-\bar{\omega}_{s_1})T}\right.\notag\\
&\left.\times\left[\frac{1}{(\bar{\omega}_{s_1}-\omega_p-\omega_-+i\gamma_a)(\bar{\omega}_{s_2}-\bar{\omega}_{s_1}-2i\gamma_a)}-\frac{2i\delta}{k+2i\delta}\frac{1}{\bar{\omega}_{s_2}-\bar{\omega}_{s_1}-i(2\gamma_a+k)}\right.\right.\notag\\
&\left.\left.\times\left(\frac{1}{\bar{\omega}_{s_1}-\omega_p-\omega_-+i\gamma_a}-\frac{1}{\bar{\omega}_{s_1}-\omega_p-\omega_++i(\gamma_a+k)}\right)\right]\right)+(s_1\leftrightarrow s_2).
\end{align}
Note, that Eq. (\ref{eq:TSJ21}) has zero time resolution as the dependence on $T$ does not depend upon matter parameters. This is caused by two photon detection events in detector $s$ with narrow frequency gating. To restore time resolution one can use only single detection event e.g. $\bar{\omega}_{s_1}$ and integrate over $\bar{\omega}_{s_2}$. Neglecting the background terms that have no resonant features we obtain
\begin{align}\label{eq:TSJ21i}
&S_{IFSRS}^{(2,1)}(\bar{\omega}_{s_1},\bar{\omega}_r;\omega_p,T)=-\mathcal{I}\frac{1}{\hbar^4}|\mathcal{E}_p|^2|\mathcal{E}_a|^2\mathcal{D}^2(\omega_p)\sum_{a,c}\alpha_{ac}^2|\mu_{ag}|^2e^{-2\gamma_aT}\left(R_q^{(2,1)}(\bar{\omega}_s,\bar{\omega}_r,2\gamma_a,\omega_--i\gamma_a)\right.\notag\\
&\left.-\frac{2i\delta e^{-kT}}{k+2i\delta}[R_q^{(2,1)}(\bar{\omega}_s,\bar{\omega}_r,2\gamma_a+k,\omega_--i\gamma_a)-R_q^{(2,1)}(\bar{\omega}_s,\bar{\omega}_r,2\gamma_a+k,\omega_+-i(\gamma_a+k)]\right),
\end{align}
where
\begin{align}\label{eq:Rq21}
R_q^{(2,1)}(\bar{\omega}_s,\bar{\omega}_r,\gamma,\Omega)=\frac{\Phi^{*}(\bar{\omega}_s,\bar{\omega}_r)\Phi(\bar{\omega}_s+i\gamma,\bar{\omega}_r)}{\bar{\omega}_s-\omega_p-\Omega}.
\end{align}

Finally, for the time-gated $S_{IFSRS}^{(1,1)}$ signal (\ref{eq:S11a0}) we obtain
\begin{align}
&S_{IFSRS}^{(1,1)a}(\bar{t}_s,\bar{\omega}_r;\omega_p,T)=\mathcal{R}\frac{1}{\hbar^4}\mathcal{D}(\omega_p)|\mathcal{E}_a|^2|\mathcal{E}_p|^2\theta(\bar{t}_s-T)\sum_{a,c}|\mu_{ag}|^2\alpha_{ac}^2\Phi^{*}(\bar{t}_s,\bar{\omega}_r)e^{-2\gamma_a\bar{t}_s}\notag\\
&\times\left[\Phi(\omega_p-\omega_++i\gamma_a,\bar{\omega}_r)e^{[i(\omega_+-\omega_p)+\gamma_a](\bar{t}_s-T)}-\frac{2i\delta}{k+2i\delta}e^{-k\bar{t}_s}\right.\notag\\
&\left.\times\left(\Phi(\omega_p-\omega_++i(\gamma_a+k),\bar{\omega}_r)e^{[i(\omega_+-\omega_p)+k+\gamma_a](\bar{t}_s-T)}-\Phi(\omega_p-\omega_-+i\gamma_a,\bar{\omega}_r)e^{[i(\omega_--\omega_p)+\gamma_a](\bar{t}_s-T)}\right)\right],
\end{align}
where we assume a broadband entangled light and $\sigma_0\gg \gamma_a, k$ and neglected the residues of the two photon amplitude. Similarly  for (\ref{eq:S11b0}) we obtain
\begin{align}
&S_{IFSRS}^{(1,1)b}(\bar{t}_s,\bar{\omega}_r;\omega_p,T)=\mathcal{R}\frac{1}{\hbar^4}\mathcal{D}(\omega_p)|\mathcal{E}_a|^2|\mathcal{E}_p|^2\theta(\bar{t}_s-T)\sum_{a,c}|\mu_{ag}|^2\alpha_{ac}^2|\Phi(\bar{t}_s,\bar{\omega}_r)|^2\frac{1-e^{-2\gamma_a\bar{t}_s}}{2\gamma_a}.
\end{align}
The frequency gated signal (\ref{eq:S11a1}) for the TSJ model is
\begin{align}\label{eq:TSJ11}
&S_{IFSRS}^{(1,1)a}(\bar{\omega}_s,\bar{\omega}_r;\omega_p,T)=-\mathcal{I}\frac{1}{\hbar^4}|\mathcal{E}_p|^2|\mathcal{E}_a|^2\mathcal{D}(\omega_p)\sum_{a,c}\alpha_{ac}^2|\mu_{ag}|^2e^{-2\gamma_aT}\left(R_q^{(2,1)}(\bar{\omega}_s,\bar{\omega}_r,2\gamma_a,-\omega_+-i\gamma_a)\right.\notag\\
&\left.-\frac{2i\delta e^{-kT}}{k+2i\delta}[R_q^{(2,1)}(\bar{\omega}_s,\bar{\omega}_r,2\gamma_a+k,-\omega_+-i\gamma_a)-R_q^{(2,1)}(\bar{\omega}_s,\bar{\omega}_r,2\gamma_a+k,-\omega_--i(\gamma_a+k)]\right),
\end{align}
where
\begin{align}\label{eq:Rq11}
R_q^{(1,1)}(\bar{\omega}_s,\bar{\omega}_r,\gamma,\Omega)=\frac{\Phi^{*}(\bar{\omega}_s,\bar{\omega}_r)\Phi(\omega_p+\Omega-i\gamma,\bar{\omega}_r)}{\bar{\omega}_s-\omega_p-\Omega}.
\end{align}
Similarly Eq. (\ref{eq:S11b1}) gives
\begin{align}
S_{IFSRS}^{(1,1)b}(\bar{\omega}_s,\bar{\omega}_r;\omega_p,T)&=\mathcal{R}\frac{1}{\hbar^4}\mathcal{D}(\omega_p)|\mathcal{E}_a|^2|\mathcal{E}_p|^2\sum_{a,c}|\mu_{ag}|^2\alpha_{ac}^2\Phi^{*}(\bar{\omega}_s,\bar{\omega}_r)\frac{1}{2\gamma_a}\notag\\
&\times\left[\Phi(\bar{\omega}_s,\bar{\omega}_r)-e^{-2\gamma_aT}\Phi(\bar{\omega}_s+2i\gamma_a,\bar{\omega}_r)\right].
\end{align}
The latter is a background term and has no resonant features that give Raman resonances. We therefore neglect it in the simulations.

Summarizing Eqs. (\ref{eq:Sc3}), (\ref{eq:TSJ01}), (\ref{eq:TSJ21i}), and (\ref{eq:TSJ11}) we obtain Eq. (\ref{eq:Sq3}) with gated Raman responses given by Eqs. (\ref{eq:Rc}), (\ref{eq:Rq01}), (\ref{eq:Rq21}), and (\ref{eq:Rq11}).

\section{IFSRS with separable correlated state}\label{sec:sep}

The general two-photon state in Eq. (\ref{eq:twin}) can be described by the density matrix
\begin{align}\label{eq:rho0}
\rho_0=\int_{-\infty}^{\infty}d\omega_sd\omega_s'd\omega_rd\omega_r'\Phi^{*}(\omega_s',\omega_r')\Phi(\omega_s,\omega_r)|1_{\omega_s},1_{\omega_r}\rangle\langle1_{\omega_s'}1_{\omega_r'}|.
\end{align}
One can further construct a different state with the same mean energy and the same single-photon spectrum, and thus would give the same single-photon transition probability. For instance if we take a diagonal part of the density matrix in Eq. (\ref{eq:rho0})
\begin{align}\label{eq:rho1}
\rho_0=\int_{-\infty}^{\infty}d\omega_sd\omega_r|\Phi(\omega_s,\omega_r)|^2|1_{\omega_s},1_{\omega_r}\rangle\langle1_{\omega_s}1_{\omega_r}|,
\end{align}
which results from disentanglement of the entangled state. Using this density operator one can compute the four-point correlation function of the electric field that enters all IFSRS signals:
\begin{align}\label{eq:4ptsep}
\langle E_s^{\dagger}(\omega_1)E_r^{\dagger}(\omega_2)E_r(\omega_3)E_s(\omega_4)\rangle=|\Phi(\omega_1,\omega_2)|^2\delta(\omega_1-\omega_4)\delta(\omega_2-\omega_3).
\end{align}
Using Eq. (\ref{eq:4ptsep}) the $S_{IFSRS}^{(0,1)}$ signal (\ref{eq:TSJ01}) yields
\begin{align}\label{eq:TSJ01sep}
&S_{IFSRS}^{(0,1)sep}(\bar{\omega}_s,\bar{\omega}_r;\omega_p,T)=-\mathcal{I}\frac{1}{\hbar^4}|\mathcal{E}_p|^2|\mathcal{E}_a|^2\mathcal{D}^2(\omega_p)\delta(0)\sum_{a,c}\alpha_{ac}^2|\mu_{ag}|^2\int_{-\infty}^{\infty}\frac{d\omega}{2\pi}|\Phi(\omega,\bar{\omega}_r)|^2\notag\\
&\left(\frac{1}{2\gamma_a[\omega-\omega_-+i\gamma_a]}-\frac{2i\delta}{[k+2i\delta][2\gamma_a+k]}\left[\frac{1}{\omega-\omega_-+i\gamma_a}-\frac{1}{\omega-\omega_++i(\gamma_a+k)}\right]\right).
\end{align}
where $\delta(0)\simeq 1/\gamma$ is a delta-function of zero argument. The $S_{IFSRS}^{(1,1)}$ signal (\ref{eq:TSJ11}) then reads
\begin{align}\label{eq:TSJ11sep}
&S_{IFSRS}^{(1,1)sep}(\bar{\omega}_s,\bar{\omega}_r;\omega_p,T)=\mathcal{I}\frac{1}{\hbar^4}|\mathcal{E}_p|^2|\mathcal{E}_a|^2\mathcal{D}(\omega_p)\delta(0)\sum_{a,c}\alpha_{ac}^2|\mu_{ag}|^2|\Phi(\bar{\omega}_s,\bar{\omega}_r)|^2\notag\\
&\left(\frac{1}{2\gamma_a[\bar{\omega}_s+\omega_++i\gamma_a]}-\frac{2i\delta}{[k+2i\delta][2\gamma_a+k]}\left[\frac{1}{\bar{\omega}_s+\omega_++i\gamma_a}-\frac{1}{\bar{\omega}_s+\omega_-+i(\gamma_a+k)}\right]\right),
\end{align}
Similarly  $S_{IFSRS}^{(2,1)}$ signal Eq. (\ref{eq:TSJ21}) becomes
\begin{align}\label{eq:TSJ21sep}
&S_{IFSRS}^{(2,1)sep}(\bar{\omega}_s,\bar{\omega}_r;\omega_p,T)=-\mathcal{I}\frac{1}{\hbar^4}|\mathcal{E}_p|^2|\mathcal{E}_a|^2\mathcal{D}^2(\omega_p)\delta(0)\sum_{a,c}\alpha_{ac}^2|\mu_{ag}|^2|\Phi(\bar{\omega}_s,\bar{\omega}_r)|^2\notag\\
&\left(\frac{1}{2\gamma_a[\bar{\omega}_s-\omega_-+i\gamma_a]}-\frac{2i\delta}{[k+2i\delta][2\gamma_a+k]}\left[\frac{1}{\bar{\omega}_s-\omega_-+i\gamma_a}-\frac{1}{\bar{\omega}_s-\omega_++i(\gamma_a+k)}\right]\right).
\end{align}

\newpage


\providecommand{\latin}[1]{#1}
\providecommand*\mcitethebibliography{\thebibliography}
\csname @ifundefined\endcsname{endmcitethebibliography}
  {\let\endmcitethebibliography\endthebibliography}{}

 \begin{figure*}[h]
\begin{center}
\includegraphics[trim=0cm 0cm 0cm 0cm,angle=0, width=0.95\textwidth]{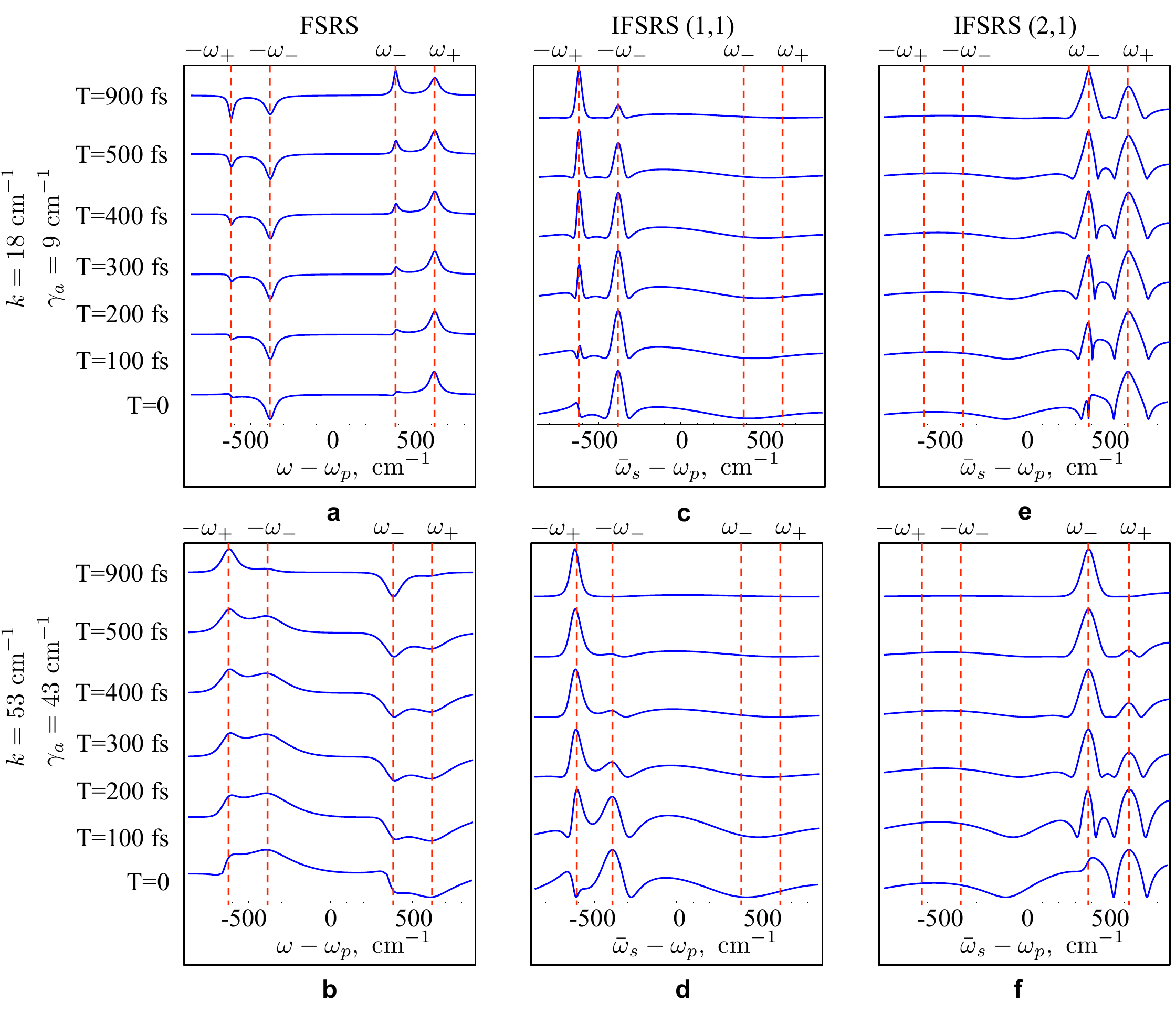}
\end{center}
\caption{(Color online) First column: $\textbf{a}$ - series of the snapshots using classical FSRS signal (\ref{eq:Sc3}) for a time evolving vibrational mode i vs $\omega-\omega_p$ for slow tunneling rate $k=18$ cm$^{-1}$ and narrow dephasing $\gamma_a=9$ cm$^{-1}$, $\textbf{b}$ - same as $\textbf{a}$ but for fast tunneling rate $k=53$ cm$^{-1}$ and broad dephasing $\gamma_a=43$ cm$^{-1}$. Second column: $\textbf{c}$, $\textbf{d}$ - same as $\textbf{a}$, $\textbf{b}$ but for for  $S_{IFSRS}^{(1,1)}$ in Eq. (\ref{eq:TSJ11}) . Third column: $\textbf{e}$ and $\textbf{f}$ - same as $\textbf{a}$, $\textbf{b}$ but for $S_{IFSRS}^{(2,1)}$ given by Eq. (\ref{eq:TSJ21i}) vs $\bar{\omega}_{s}-\omega_p$. All parameters are the same as in Fig. \ref{fig:tsj}.
}
\label{fig:tsjsup}
\end{figure*}

 \begin{figure*}[h]
\begin{center}
\includegraphics[trim=0cm 0cm 0cm 0cm,angle=0, width=0.65\textwidth]{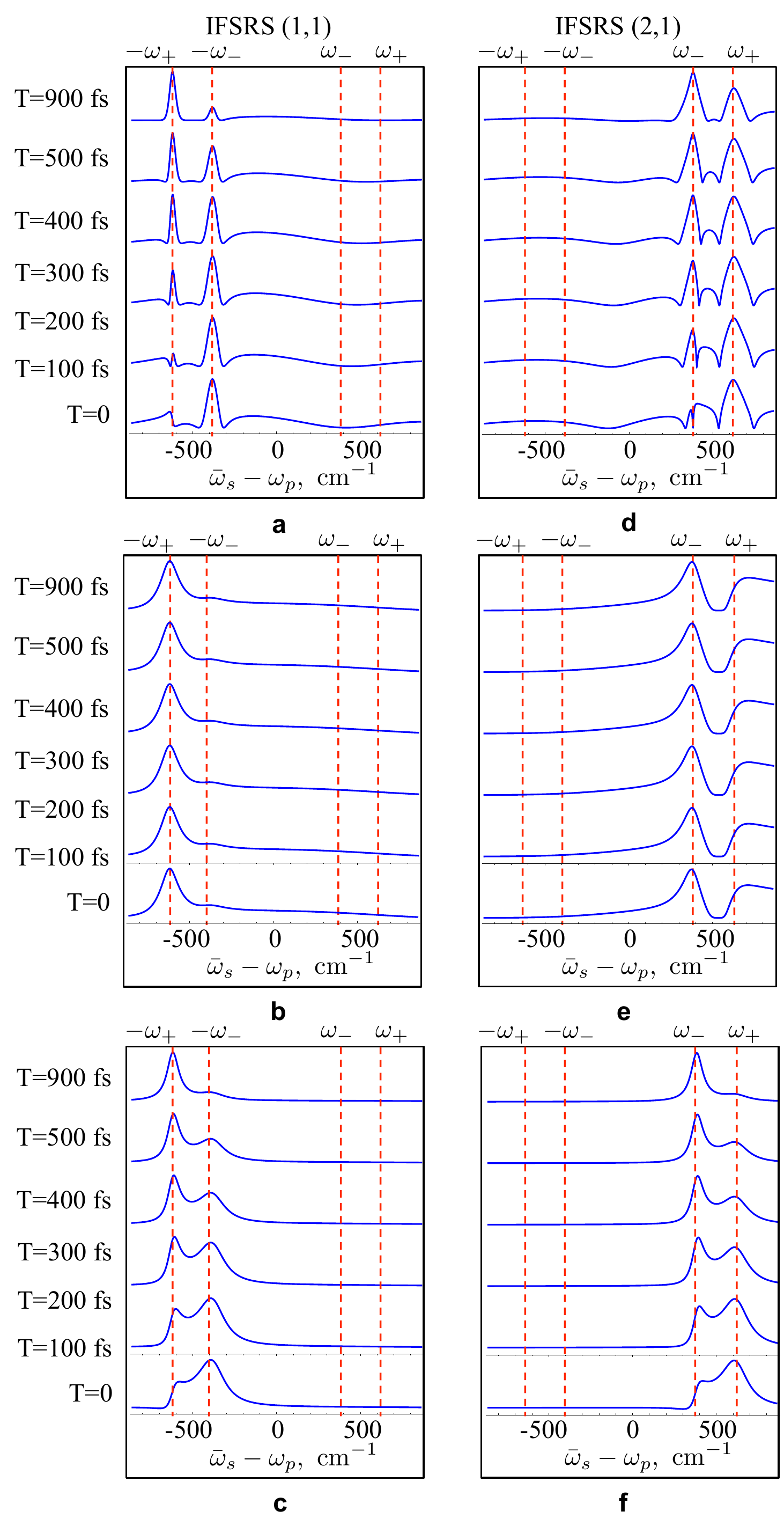}
\end{center}
\caption{(Color online) Left column: Series of the snapshots using $S_{IFSRS}^{(1,1)}$ signal (\ref{eq:TSJ11})  vs $\bar{\omega}_s-\omega_p$ for entangled state (\ref{eq:twin1}) - $\mathbf{a}$, correlated - $\mathbf{b}$ and uncorrelated - $\mathbf{c}$ separable states. $\mathbf{d}$, $\mathbf{e}$, and $\mathbf{f}$ - same as $\mathbf{a}$, $\mathbf{b}$, and $\mathbf{c}$ but for $S_{IFSRS}^{(2,1)}$ signal (\ref{eq:TSJ21i}). All parameters are the same as in Fig. \ref{fig:tsj}.
}
\label{fig:sepsup}
\end{figure*}

\end{document}